\RequirePackage{fix-cm}
\documentclass[twocolumn]{svjour3}          
\smartqed  
\usepackage{graphicx}
 \journalname{Theor Ecol}
\begin{document}
\bibliographystyle{plainnat}

\title{Phytoplankton competition in deep biomass maximum}


\author{Alexei B. Ryabov}


\institute{A.B. Ryabov \at
              ICBM, University of Oldenburg, 26111 Oldenburg, Germany \\
              Tel.: +49-441-798-3638\\
              Fax: +49-441-798-3404\\
              \email{a.ryabov@icbm.de}           
}

\date{Received: 10 January 2011 / Accepted: 4 January 2012}

\maketitle

\begin{abstract}
Resource competition in heterogeneous environments is still an unresolved  problem of theoretical ecology. In this article I analyze competition between two phytoplankton species in a deep water column, where the distributions of main resources (light and a limiting nutrient) have opposing gradients and co-limitation by both resources causes a deep biomass maximum.  Assuming that the species have a trade-off in resource requirements and the water column is weakly mixed, I apply the invasion threshold analysis (Ryabov and Blasius 2011) to determine relations between environmental conditions and phytoplankton composition. Although species deplete resources in the interior of the water column, the resource levels at the bottom and surface remain high.  As a result, the slope of resources gradients becomes a new crucial factor which, rather than the local resource values, determines the outcome of competition. 
The value of resource gradients nonlinearly depend on the density of consumers.  This leads to complex relationships between environmental parameters and species composition.
In particular,  it is shown that an increase of both the incident light intensity  or bottom nutrient concentrations favors the best light competitors, while an increase of the turbulent mixing or background turbidity favors the best nutrient competitors. These results might be important for prediction of species composition in deep ocean.
\end{abstract}

\section{Introduction}
Primary production forms the basis of metabolic activity of the ocean. Distinct phytoplankton groups contribute differently in the sequestration of ${\rm CO_2}$ (Frankignoulle et al. 1994; Smetacek 1999), production of oxygen (Falkowski and Isozaki 2008), support of marine food webs (Christoffersen 1996), etc. 
A shift in the species composition may dramatically affect functioning of the whole ecosystem (Walther et al. 2002; Cerme\~{n}o et al. 2008; Paerl and Huisman 2009).
However, in spite of the principal role of resource competition in the community structuring, the conditions of coexistence and competitive exclusion in spatially variable environments still remain largely unknown.

The classical theory, advanced by MacArthur (1972), Le\'{o}n and Tumpson (1975), and Tilman (1980, 1982), analyses resource competition in uniform environments and shows that stationary coexistence of two species on two resources is possible only if 
growth of each species is finally restricted by its most limiting resource. 
The same results hold for competition in a mixed water column where light exponentially decreases with depth (Huisman and Weissing 1995; Diehl 2002). However, in weakly-mixed systems these conditions may not be met. Competitors can be simultaneously limited by two or more resources, if their favorable habitats are surrounded by areas lacking these resources.

For instance, in deep oligotrophic aquatic systems  the light intensity reduces with depth, whereas concentrations of nutrients typically have opposing gradients. As a consequence, species with  distinct resource requirements can have maxima of production at different depths, which potentially decreases niche overlaps and increases biodiversity  (Chesson 1990, 2000). However, a general extension of the competition theory to continuous spatially variable habitats leads to difficult mathematical problems (Grover 1997) and was addressed mainly in the mathematical literature (Hsu and Waltman 1993). The analysis of this problem from the ecological point of view (Huisman et al. 1999; Yoshiyama et al. 2009; Dutkiewicz et al. 2009; Ryabov et al. 2010; Ryabov and Blasius 2011) is still far from complete and further research which would connect results for uniform and spatially variable systems is required. 

In this article I analyze competition between two phytoplankton species in a deep weakly-mixed water column,
assuming that limitation by light in deep layers and limitation by nutrients at shallow depths
cause deep chlorophyll or biomass maxima (Holm-Hansen and Hewes 2004; Kononen et al. 2003), which are a wide spread phenomenon in oligotrophic basins (Abbott et al. 1984; Karl and Letelier 2008). The location of a favorable layer in such systems is not fixed, rather it depends on initial and boundary conditions, the stage of the relaxation process, etc. (Klausmeier and Litchman 2001; Yoshiyama and Nakajima 2002; Ryabov et al. 2010). Furthermore, a species, establishing at a certain depth, changes resource distributions and may affect all other species throughout the water column. Thereby this species acts as an ecosystem engineer, modifying its nutrient and light environment.

For the analysis of competition in such a system Ryabov and Blasius (2011) recently introduced the notion of an invasion threshold, defined as a line (in case of two limiting resources) or a hypersurface (in general) in space of resource requirements, which separates the species that can grow in the presence of a resident species from those that cannot grow.
The form and location of invasion thresholds depend on the characteristics of competitors as well as on the environmental conditions. The investigation of these dependences in the phytoplankton model reveals conditions that favor different competitors and can explain shifts in the species composition caused by environmental changes.

\section{Model}

\begin{figure*}[tb]
\begin{center}
\centerline{\includegraphics[width=13cm]{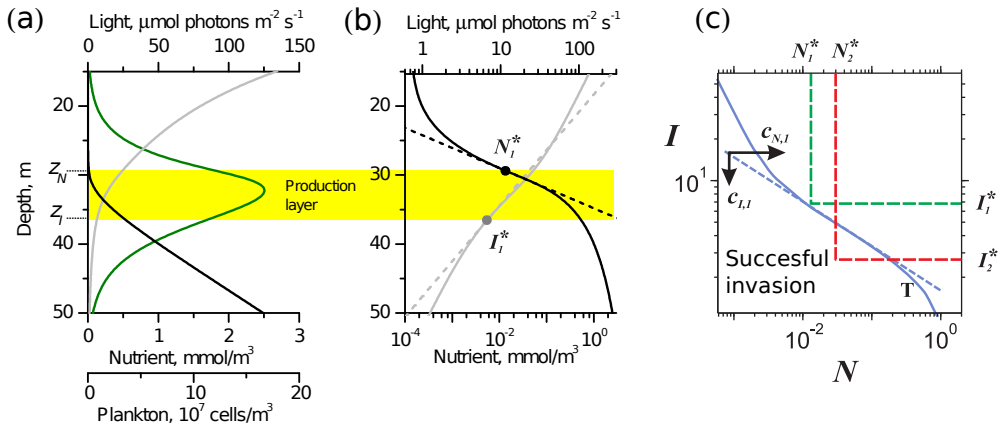}}
\caption{(a) Distributions of the nutrient concentration (black), light intensity (gray), and phytoplankton biomass (green) in the model Eqn.~\ref{eq:1sppl}-\ref{eq:grrate}. 
(b) Distributions of the nutrient and light (solid lines) and exponential fitting (dashed lines) according to  Eqn.~\ref{eq:explight} and \ref{eq:expnutrient} plotted in the logarithmic scale. The yellow shaded area is the production layer, where  the growth rate excesses mortality since $N>N^*$ and $I>I^*$.
(c) Zero net growth isoclines of species 1 and species 2 (green and red dashed lines, respectively); invasion threshold (blue lines) in the presence of species 1 as the resident under the conditions listed in Table~\ref{tab:1}.
The blue solid line was calculated numerically, by test of more than 5000 invaders with different half-saturation constants; the blue dashed line is a first order approximation with slope $\gamma_1=c_{I,1}/c_{N,1}$ in the log-log scale (Eq.~\ref{eq:criteria})
}
\label{fig:model}
\end{center}
\end{figure*}

Competition between two phytoplankton species for light and a limiting nutrient (e.g., nitrogen or phosphorus) in a water column can be modeled in terms of a nonlocal system of coupled reaction-diffusion equations (Radach and Maier-Reimer 1975; Jamart et al. 1977; Klausmeier and Litchman 2001; Huisman et al. 2006) 
\begin{eqnarray}
\frac{\partial P_i}{\partial t} &=&
  \mu_i(N, I) P_i - m_i P_i + D\frac{\partial^2 P_i}{\partial z^2} \ ,
\label{eq:1sppl}\\
 \nonumber \\
\frac{\partial N}{\partial t} &=&
 - \sum_{i=1}^n \alpha_i \mu_i(N, I) P_i + 
D\frac{\partial^2 N}{\partial z^2}  \label{eq:1spntr} \ ,
\end{eqnarray}
where $P_i(z, t)$ is the population density of the phytoplankton species $i$ at depth $z$ and time $t$, $\mu_i(I, N)$ is the growth rate, which depends on the local values of the light intensity, $I(z, t)$, and nutrient concentration, $N(z, t)$, $m_i$ is the mortality rate, $D$ is the turbulent diffusivity, and $\alpha_i$ is the nutrient content of a phytoplankton cell. 

The light intensity decreases with depth owing to the absorption of light by water and phytoplankton biomass (Kirk 1994)
\begin{equation}
I(z) = I_{in} \exp \left[ -K_{bg} z - \int^z_0 \sum_{i=1}^n k_i P_i(\xi, t) d \xi \right] \ ,
\label{eq:1splight}
\end{equation}
where $I_{in}$ is the intensity of incident light, $K_{bg}$ is the water turbidity and $k_i$ is the attenuation coefficient of phytoplankton cells.

Assume that both resources are essential (von Liebig's law of minimum) and the resource limitation of growth can be parametrized by the Monod kinetics (Turpin 1988), then the growth rate of species $i$ follows
\begin{equation}
\mu_i(N, I) =  \mu_{\max, i} \min \left\{
\frac{N}{H_{N,i} + N}, \frac{I}{H_{I,i} + I}  \right\} \ , 
\label{eq:grrate}
\end{equation}
where $\mu_{\max, i}$ is the maximal growth rate, and $H_{N,i}$ and $H_{I,i}$ are the half-saturation constants, which define the species resources requirements. 

The phytoplankton cells cannot diffuse across the surface and bottom of the water column
\[
\left.\frac{\partial P(z, t)}{\partial z}\right|_{z = 0} = 0 \ , \qquad
\left.\frac{\partial P(z, t)}{\partial z}\right|_{z = Z_B} = 0 \ .
\]
The surface is also impenetrable for the nutrient, and the nutrient concentration at the bottom is constant
\[
\left.\frac{\partial N(z, t)}{\partial z}\right|_{z = 0} = 0\ , \qquad N(Z_B) = N_B \ .
\]

Fig.~\ref{fig:model} shows typical equilibrium resource and biomass distributions in this model. 
The yellow shaded area indicates the production layer where both resources are available and the growth rate excesses mortality, $\mu(N, I) > m$.
In a non-uniform system the total net growth on a favorable area should be large enough to compensate for losses into adjacent unfavorable areas (Ryabov and Blasius 2008).  Denote the depths, which confine this area, as $z_N$ if $N$ limits species growth and $z_I$ if light limits species growth. The resource availability reaches at these depths the critical values
\begin{equation}
  N^*_i =  \frac{m_i}{\mu_{max, i}-m_i}  H_{N, i}   \ , \  I^*_i = \frac{m_i}{\mu_{max, i}-m_i} H_{I, i} \ ,
\label{eq:critical} 
\end{equation}
at which growth equals  mortality.
The location of the production layer depends on the critical resource values and environmental parameters.
In the present  article, to focus on the influence of resource gradients, the parameters are chosen to reproduce deep biomass maxima: the production layer is located in deep layers and the biomass density vanishes at the bottom and surface (see Table~\ref{tab:1} for parameters). 

The invasion analysis is applied to determine outcomes of competition.
Initially one species (the resident) grows alone during a sufficiently long time and then another species (the invader) can grow from a small biomass density. Two species coexist if each of them can invade in the presence of its competitor. For the given model parameters  the distributions of biomass and nutrients approach to equilibrium after approximately 500 simulation days. To make sure that the resident is at equilibrium this time interval is increased up to 2000 days, after which an invader can grow and the system is simulated for further 18000 days to obtain the final competition outcome. 

The initial distribution of nutrients change linearly from 0 at the surface to $N_B$ at the bottom. The phytoplankton species have initially uniform distribution of small density, $P = 100$ cells/m$^3$.
For the numerical solution the partial differential equations were discretized on a grid of 0.25 m. The resulting system of ordinary differential equations was solved by the CVODE package (http://www.netlib.org/ode) using the backward differentiation method.

\section{Competition in a spatially variable environment} 
Assume that the competitors trade off in resource requirements: species 1 is a better nutrient competitor and species 2 is a better competitor for light (Fig.~\ref{fig:model}c),
and consider the invasion of species 2, assuming that species 1 has reached an equilibrium distribution.

\subsection{Invasion threshold}
Although the resident species depletes resources in the middle of the water column, the light intensity at the surface and nutrient concentrations at the bottom are still high (Fig.~\ref{fig:model}a, b). Due to a difference in resource requirements the maximum of production of species 2 can be shifted towards, for instance, the nutrient supply, and even a strong requirement for high nutrient concentrations can be compensated by adaptation to a low light intensity and {\it vice versa}. As a result, the invasion threshold takes the shape of a curve (the blue solid line in Fig.~\ref{fig:model}c is calculated by invasion of approximately 5000 species with different half-saturation constants).

The shape of invasion thresholds depends on the resource distribution. However, to find this dependence we should find the principal eigenvalues corresponding to the boundary problem of Eqs.~\ref{eq:1sppl}--\ref{eq:1spntr} (Hsu and Waltman 1993; Grover 1997; Ryabov and Blasius 2011), which can be solved in general only numerically (Troost et al. 2005). However, a simple analysis can be performed for competitors with close resource requirements. Then the maximum production of invaders occurs in the vicinity of the resident production layer (Fig.~\ref{fig:model}c) and it is possible to express the growth rate of invaders in terms of the growth rate of the resident.

    \subsection{Resource distribution}
Due to the diffusion of cells an essential part of the resident biomass is located outside the production layer. Consequently the resources below and above this layer are depleted to values which are even smaller than the critical values of the resident $N^*_1$ and $I^*_1$. For instance, the nutrient concentration above the biomass maximum can be few orders of magnitude smaller than $N^*_1$ (Fig.\ref{fig:model}b). Therefore, the resident species shapes resource gradients both within and outside of its production layer. 

The shape of resource distributions in this area will play a key role for the further analysis.
It is commonly assumed, that the light intensity is exponentially distributed, while nutrient concentrations can  be fitted by a line (see the gray and black line, respectively, in Fig.~\ref{fig:model}a). However, 
the linear part of the nutrient profile is typically observed in the deep layers below the biomass maximum, 
where the light rather than the nutrient limitation determines species growth. Similarly, above the production layer the growth is nutrient limited and the net growth rate is negative independently of the shape of the nutrient distribution. The shape of resource distributions is crucial within the production layer, where the local net growth rate strongly correlates with local resource availability. As shown in Fig.\ref{fig:model}b, in this area both the light and nutrient distributions closely follow exponential dependences (straight dashed lines on the logarithmic scale). 

The exponential shape of resource distributions is a crucial point of the following theory. For this reason, it is important to note that this shape is not just an artifact of a specific model, but it was also found in field observations. For instance, Karl and Letelier (2008) clearly demonstrate exponential dependence of nutrient concentrations in the area of nutrient consumption. 
The emergence of such distributions can have different nature. In particular, it is easy to show that it emerges, when the biomass variation within the production layer is small. 

Suppose that the phytoplankton density can be approximated by a rectangular  distribution with the constant density, 
$$P_0=\frac{1}{z_I - z_N} \int_{z_N}^{z_I} P(z) dz$$
 within and in the vicinity of the production layer. This implies a small mortality level outside the production layer, $m \ll \mu_{max}$, so that the biomass can diffuse without essential losses. Then, according to Eq.~\ref{eq:1splight}, the absolute value of the logarithmic gradient of light intensity is constant and equals
\begin{equation}
c_I = -\frac{d \ln I(z)}{dz} = K_{bg} + k P_0 .
\label{eq:ci}
\end{equation}
Thus, the light distribution can be approximated as
\begin{equation}
\widetilde{I}(z)  = I^* e^{- c_{I}\,  (z - z_{I})} \ .
\label{eq:explight}
\end{equation}

To find the nutrient distribution note that according to Eqn.~\ref{eq:critical} in the limit of small mortality the critical nutrient concentrations are also small, $N^* \ll H_N$, therefore the growth rate close to the depth $z_N$ can be linearized as 
$
\mu_N(N) \approx \mu_{max} N/H_N \, . 
$
Substituting this expression into Eq.~\ref{eq:1spntr} we obtain in equilibrium 
$$
-\alpha \mu_{max} \frac{N}{H_N} P_0 + D \frac{d^2 N}{d z^2} = 0 \ .
$$
A solution to this equation, that monotonically increases with depth, gives the equilibrium nutrient distribution in the vicinity of the depth  $z_{N}$,
\begin{equation}
\widetilde{N}(z)  = N^*  e^{c_{N,1} \, (z - z_{N})} 
\label{eq:expnutrient}
\end{equation}
with the logarithmic gradient
\begin{equation}
c_N = \frac{d \ln N(z)}{dz} = \sqrt{\frac{\alpha \mu_{max} P_0}{D H_N}} \ .
\label{eq:cn}
\end{equation}

Thus in the vicinity of the production layer both the light and nutrient distribution have exponential shape, which implies that the logarithmic resource gradients have small variations with depth.

\subsection{Approximate calculation of the invasion thresholds}

\begin{figure*}[tb]
\begin{center}
\centerline{
\includegraphics[width=0.9\textwidth]{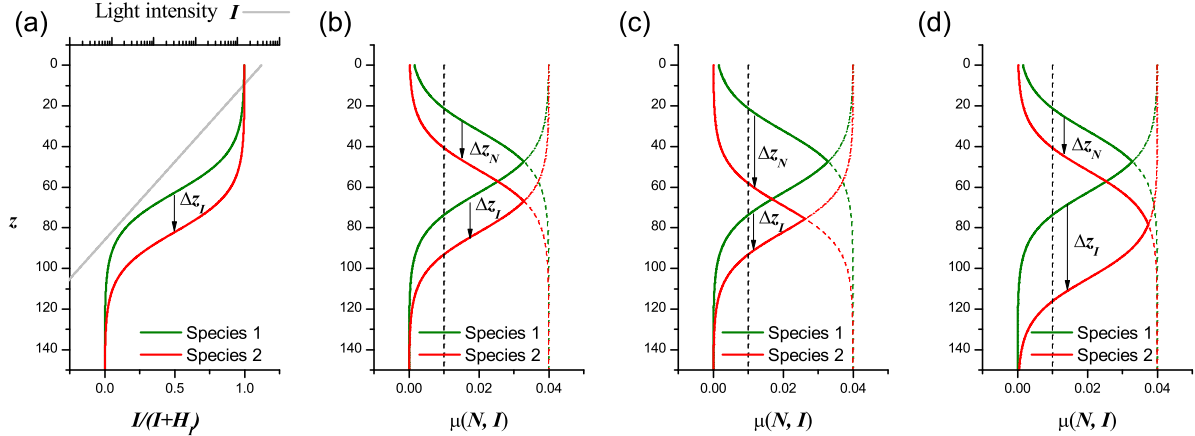}}
\caption{Schematic profiles of resource limitation of species growth, assuming that resource distributions are exponentially shaped.
(a) If $I(z) \propto e^{-c_I z}$ (gray straight line on a logarithmic scale) then a difference in half-saturation constants of the species results in a parallel translation of the Monod function by $\Delta z_I = -c_I^{-1} \ln \left( H_{I, 2}/H_{I, 1} \right)$  along z-axis.
In Figs.b-d solid lines show growth rates of each species, $\mu(N, I)$. The limitations of growth by particular resources, $\mu_N(N)$ and $\mu_I(I)$, are shown as dashed and dot-dashed lines respectively in the regions where they do not coincide with $\mu(N, I)$.  The black dashed line shows the mortality rate, $m$.
(b)  $\Delta z_I = \Delta z_N$, parallel translation of the growth rate profiles, leads to the same total net growth of the species.
In (c) and (d) either $\Delta z_I < \Delta z_N$ or $\Delta z_I > \Delta z_N$  and the net total growth rate of species  2 is either smaller or larger, respectively,  than the total net growth of species 1.
}
  \label{fig:GrShifts}
\end{center}
\end{figure*}

To gain an insight into the dependence of invasion thresholds on resource distributions 
assume that the resident (species 1) and invader (species 2) differ only in their half-saturation constants, $H_{N,i}$ and $H_{I,i}$, but are otherwise identical, i.e. $\mu_{max, 1} = \mu_{max, 2}$, $m_1 = m_2$,  and $D_1 = D_2$, (see Ryabov and Blasius (2011) for a general approach). 

An invader of small initial density has a vanishing influence on the resources, therefore
the possibility of invasion depends only on its growth rate 
in the resource distribution shaped by the resident,
$\mu_2(\widetilde{N}_1(z), \widetilde{I}_1(z))$. Consider the difference between the nutrient limitations of the resident and invader.	 
The Monod kinetics can be presented in the form 
\begin{equation}
  \frac{\widetilde{N}_1(z)}{\widetilde{N}_1(z) + H_{N,2}} = 
  \frac{\widetilde{N}_1(z)/H_{N,2}}{(\widetilde{N}_1(z)/H_{N,2}) + 1} = \mu_N\left(\frac{\widetilde{N}_1(z)}{H_{N,2}}\right)\ ,
\nonumber
\end{equation}
which shows that the nutrient limitation of growth, $\mu_N$, depends only on the ratio $\widetilde{N}_1/H_{N,2}$.

However, the mathematical identity 
\begin{equation}
\frac{e^{c z}}{H_2}   =
\frac{ e^{c(z - \Delta z)}}{H_1} \ ,\quad \mbox{where}  \   \Delta z = \frac{1}{c} \ln \frac{H_2}{H_1}  \ , 
\label{eq:trick}
\end{equation}
shows that division of an exponential function by different constants $H_1$ and $H_2$ is equivalent to a shift $\Delta z$ in position along the $z$-axis. 
Thus, using the exponential approximation for the nutrient distribution, Eq.~\ref{eq:expnutrient}, we
obtain 
\begin{equation}
\frac{\widetilde{N_1}(z)}{H_{N,2}} =     
 \frac{\widetilde{N_1}(z - \Delta z_{N})}{H_{N,1}} \ , 
\nonumber
\end{equation}
and the nutrient limitation of invader growth can be expressed through that of the resident
\begin{eqnarray}
\mu_N\left(\frac{\widetilde{N_1}(z)}{H_{N,2}}\right) &=&     
 \mu_N\left(\frac{\widetilde{N_1}(z - \Delta z_{N})}{H_{N,1}}\right) \ , 
 \nonumber \\
 \mbox{where} \quad  	\Delta z_{N} &=& \frac{1}{c_{N, 1}} \ln \frac{H_{N, 2}}{H_{N, 1}} \ .
 \label{eq:NutrShift}
\end{eqnarray}
Therefore, the profiles of nutrient limitation of resident and invader growth have the same shape, but are shifted by $\Delta z_{N}$ with respect to each other. 
If $H_{N, 2} > H_{N, 1}$, i.e. the invader needs higher nutrient concentrations, then $\Delta z_N$ is positive and the invader nutrient limitation profile is shifted downwards. Performing the same calculations for the light limitation we obtain
\begin{eqnarray}
 \mu_I\left(\frac{\widetilde{I_1}(z)}{H_{I,2}}\right) &=&     
 \mu_I\left(\frac{\widetilde{I_1}(z - \Delta z_{I})}{H_{I,1}}\right)  \ ,
  \nonumber \\
  \mbox{where} \quad  	\Delta z_I &=& -\frac{1}{c_{I, 1}} \ln \frac{H_{I, 2}}{H_{I, 1}}   \ .
\label{eq:LightShift}
\end{eqnarray}
$\Delta z_I$ has the opposite sign, because the light intensity has an inverse gradient. Assuming that the invader is a better light competitor ($H_{I, 2} < H_{N, 1}$), we obtain that $\Delta z_I$ is positive and the profile of light limitation is shifted downwards (see Fig.~\ref{fig:GrShifts}a). 

\begin{figure*}[tb]
\begin{center}
\centerline{
\includegraphics[width=0.7\textwidth]{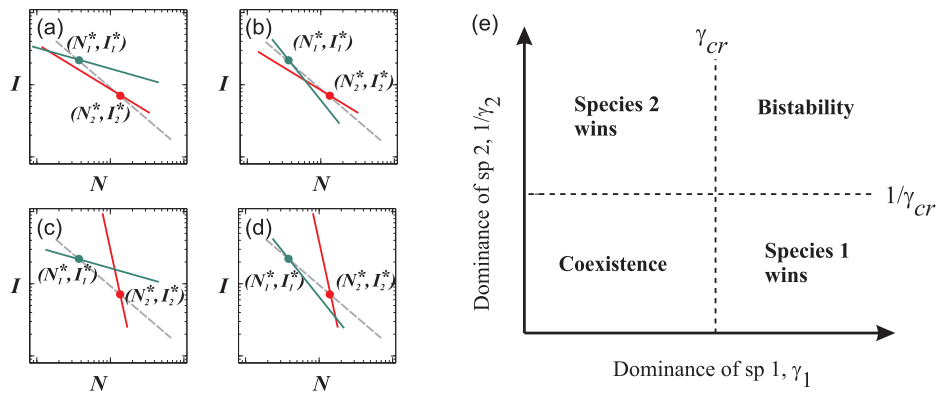}}
\caption{(Left panel) Competition outcome in dependence of the slope of invasion thresholds of species 1 (green) and species 2 (red). The gray dashed line shows the critical slope, $\gamma_{cr}$.
(a) $\gamma_{1, 2} < \gamma_{cr}$ -- species 2 wins, (b)  $\gamma_{2} < \gamma_{cr} < \gamma_{1}$ -- bistability, (c) $\gamma_{1} < \gamma_{cr} < \gamma_{2}$ -- coexistence, (d) $\gamma_{1, 2} > \gamma_{cr}$ -- species 1 wins. (Right panel, e) All outcomes can be represented in dependence on $\gamma_1$ and $1/\gamma_2$.
}
\label{fig:comp}
\end{center}
\end{figure*}

The values $\Delta z_I$ and $\Delta z_N$ show changes in the net growth arising from better adaptation to one or another resource, and the value $\Delta = \Delta z_N - \Delta z_I$ defines the difference in the size of the resident and invader's favorable habitats. If boundary effects on biomass distributions are negligible, then the fate of the invader species depends solely on the sign of $\Delta$. Indeed, if $\Delta = 0$ then the distinct resource requirements of these species result in a parallel translation of the growth rate profile along the $z$-axis (Fig.~\ref{fig:GrShifts}b). Since the resident species has zero total growth in equilibrium, the same holds for the invader and its population cannot establish. If $\Delta > 0$ then the invader habitat is even smaller, consequently the invader total net growth will be negative (Fig.~\ref{fig:GrShifts}c). By contrast, if $\Delta<0$, the invader habitat is larger and it can invade the system because its total production is positive (Fig.~\ref{fig:GrShifts}d). 

According to Eqn.~\ref{eq:critical}, when the growth and mortality rates are equal, the ratio of half-saturation constants equals the ratio of critical resource values (e.g., $H_{I, 2}/H_{I, 1} = I^*_2/I^*_1$), and the invasibility threshold can be presented in the form
\begin{equation}
 \Delta = \frac{1}{c_{N, 1}} \ln \frac{N^*_2}{N^*_1} + \frac{1}{c_{I, 1}} \ln \frac{I^*_2}{I^*_1} < 0 \ .
\label{eq:criteria}
\end{equation}
This inequality defines a first order approximation of the invasion threshold and has a straightforward geometrical interpretation: 
species 2 can invade in the presence of species 1 if its critical resource values $(N^*_2, \ I^*_2)$ 
lie below a line with slope $\gamma_1=c_{I, 1}/c_{N, 1}$ passing through the point $(N^*_1, \ I^*_1)$ in the double-logarithmic resource plane (blue dashed line in Fig.~\ref{fig:model}c) and  
the range of possible invaders is determined by the ratio of logarithmic resource gradients shaped by the resident.

The value of $\gamma$ depends on the characteristics of the resident, and species 2 growing alone will shape a distinct resource distribution with different ratio $\gamma_2 =  c_{I, 2}/c_{N, 2}$.
Without loss of generality, assume that  $N^*_2 > N^*_1$ and $I^*_2 < I^*_1$,
and denote by $\gamma_{cr}$ the slope (taken with opposite sign) of a straight
line passing through the two critical resource points (gray dashed lines in Figs.~\ref{fig:comp}a--d), 
\begin{equation}
	\gamma_{cr}=-\frac{\ln I^*_2/I^*_1}{\ln N^*_2/N^*_1} \ .
	\label{eq:gamma_cr}
\end{equation}
Then, combining Eq. \ref{eq:criteria} and its counterpart for species 2,
we obtain the outcomes of spatial resource competition.
If $\gamma_1 < \gamma_{cr}< \gamma_2$, then both species can invade the monoculture of each other and can thus coexist (Fig.~\ref{fig:comp}c). In the opposite case, $\gamma_1 > \gamma_{cr} > \gamma_2$
none of the two species can invade, which leads to alternative stable states (Fig.~\ref{fig:comp}b). Finally, one species dominates if $\gamma_{1, 2} > \gamma_{cr}$ or $\gamma_{1, 2} < \gamma_{cr}$ (Fig.~\ref{fig:comp}a, d). 

Consider in detail the conditions of coexistence. 
Fig.~\ref{fig:comp}c shows that the best nutrient competitor (species 1, green) should have a relatively shallow slope ($\gamma_1< \gamma_{cr}$) of its invasion threshold, therefore this species should shape a resource distribution with a relatively small gradient of light intensity, $c_{I, 1}$. This will provide more solar radiation for species 2 (red), which, owing to its stronger nutrient limitation, has a niche in deeper layers. At the same time the invasion threshold of species 2 should have a steeper slope ($\gamma_2 > \gamma_{cr}$), and therefore $c_{N, 2}$ should be relatively small. In other words, this species should not diminish the upward nutrient flux too much. Thus, we obtain a general rule that for coexistence each species should shape resource distributions with a relatively smaller gradient of its most limiting resource. 

It is convenient to represent all possible outcomes  of competition in dependence of  $\gamma_1$ and $1/\gamma_2$ (Fig.~\ref{fig:comp}e), which reflect the ability of a resident to shape a stronger gradient of its less limiting resource and quantify the dominance of the species over its competitor.
Then coexistence is possible if the mutual dominance of both species is small, whereas large values of $\gamma_1$ and $1/\gamma_2$ lead to bistability. 

To get another perspective on the role of resource gradients, consider invasion 
in the presence of a resident, which has shaped a large gradient of the nutrient distribution and a small gradient of light intensity, $c_{N, 1} \gg c_{I, 1}$. 
According to Eq.~\ref{eq:LightShift}, the change in the area of light limitation, $\Delta z_I$,  approaches infinity when $c_{I, 1} \rightarrow 0$, thus
even a small difference in light requirement, $\ln (I^*_{2}/I^*_{1}) < 0$, can lead to a large increase of the favorable area. Therefore, a better adaptation to this resource is very efficient. 
By contrast, a large nutrient gradient $c_{N, 1}$ makes competition much harder, because the areas of the resident and invader nutrient limitation will almost coincide ($\Delta z_N \propto 1/c_{N, 1}\rightarrow 0$,  unless $N^*_{2} \ll N^*_{1}$ or $N^*_{2} \gg N^*_{1}$, see Eq.~\ref{eq:NutrShift}). 
In the limit case $c_{N, 1} \rightarrow \infty$, competition for the nutrient becomes impossible because the habitat is virtually divided into two parts: with very low nutrient concentrations (all species are nutrient limited) and very high concentrations (the nutrient limitation plays no role). 

\begin{figure*}[tb]
\begin{center}
\centerline{\includegraphics[width=0.6\textwidth]{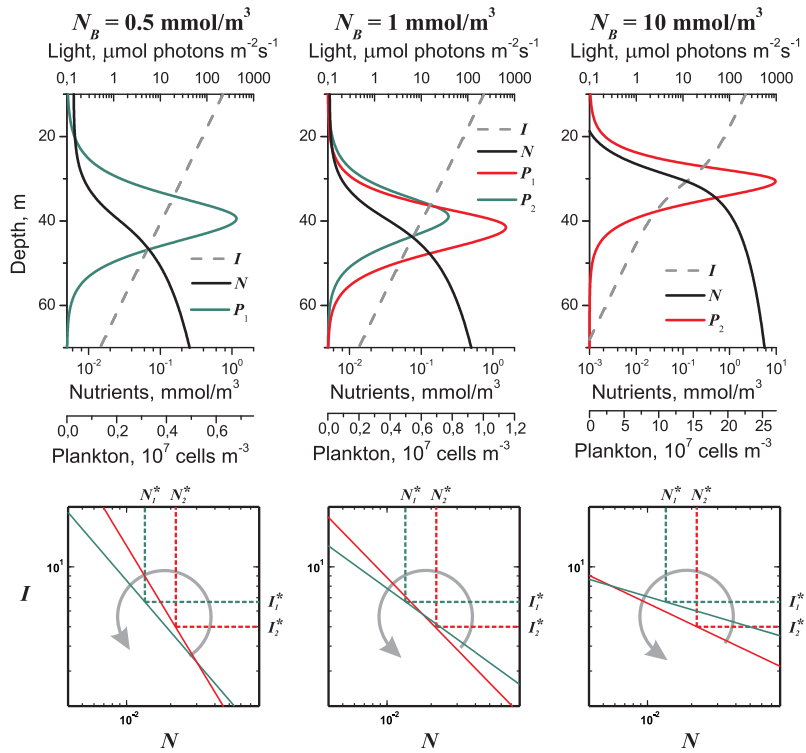}}
\caption{
Geometrical method to project the outcome of spatial resource competition in dependence of the nutrient concentration at the bottom, $N_B$.
Top panel: shift in the competition outcome between species 1 (green) and species 2 (red) in the phytoplankton model 
caused by an increase of $N_B$. Bottom panel illustrates this shift as the result of a counter-clockwise rotation of the invasion thresholds. The slopes, $\gamma_{1}$ and  $\gamma_{2}$, of the invasion thresholds in the bottom panel were calculated numerically for a monoculture of species 1 and 2. 
}
\label{fig:changeN}
\end{center}
\end{figure*} 

\section{The influence of environmental parameters on the competition outcome}
In this section the invasion threshold analysis is applied to explain shifts in the species composition predicted by model Eqn.~\ref{eq:1sppl}--\ref{eq:grrate}. First two examples demonstrate how changes in the competition outcome can be visualized as rotation of invasion thresholds and then changes in the species composition are considered in the $(N_B, I_{in})$ and $(D, K_{bg})$ planes. 

\subsection{Shifts in the species composition with $D$ and $N_B$}
\begin{figure*}[tb]
\begin{center}
\centerline{\includegraphics[width=0.6\textwidth]{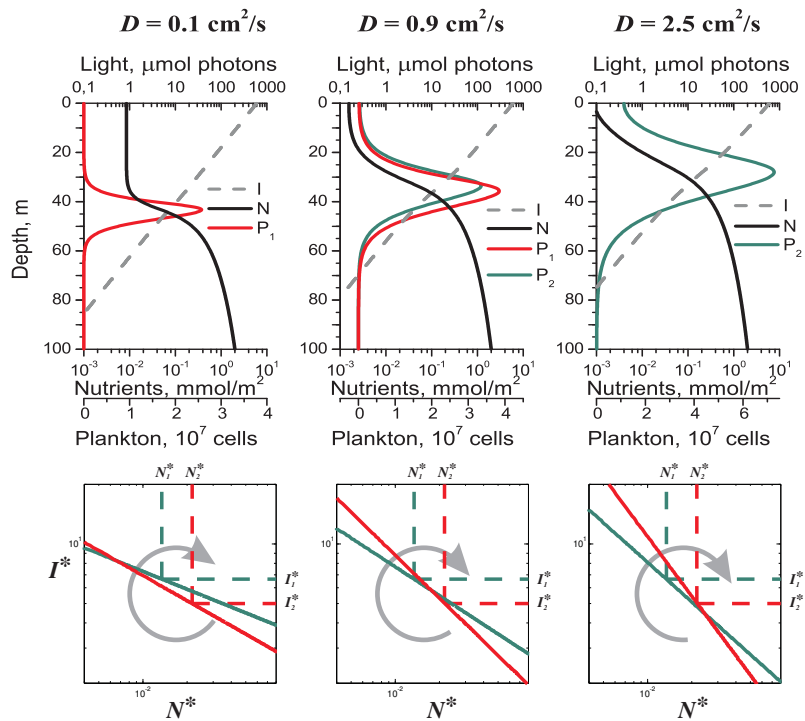}}
\caption{
The same as in Fig.~\ref{fig:changeN} but in relation to diffusivity. $N_B = 2$ mmol nutrient m$^{-3}$}
\label{fig:changeD}
\end{center}
\end{figure*}

Compare changes in the composition of phytoplankton species, caused by an increase of turbulent mixing, $D$, and bottom nutrient concentration, $N_B$.
Both these parameters increase the nutrient availability throughout the water column. In a homogeneous environment higher nutrient concentrations would provide better conditions for the better light competitor (species 2).  However, surprisingly, in the phytoplankton model these changes have opposite effect. While an increase of $N_B$ favors the better light competitor (Fig.~\ref{fig:changeN}), an increase of $D$ favors to the better nutrient competitor (Fig.~\ref{fig:changeD}). To explain these differences consider the influence of these parameters on the resource gradients and the slope of invasion thresholds. 

More nutrients at the bottom give rise to a larger and sharper biomass distribution of the resident species, which in turn yields a steeper gradient of the nutrient concentration.
As a result, with an increase of $N_B$, the low light adapted species 2 ``shades'' the nutrient flux more strongly and hinders invasion of a better nutrient competitor species 1, which, in consequence of its light limitation, occupies higher layers.  
Because a larger nutrient gradient $c_{N, i}$ leads to smaller values of $\gamma_i$, this transition can be represented as a counter-clockwise rotation of the invasion thresholds in the resource plane (the bottom panel in Fig.~\ref{fig:changeN}). 

By contrast, with an increase of mixing, $D$, the biomass distribution becomes wider and the nutrient gradient decreases (Fig.~\ref{fig:changeD}). As a result, more nutrients become available for species 1, which finally wins the competition. In the resource plane this change can be represented as a clockwise rotation of the invasion thresholds.

\begin{figure*}[tb]
\begin{center}
\centerline{\includegraphics[width=0.8\textwidth]{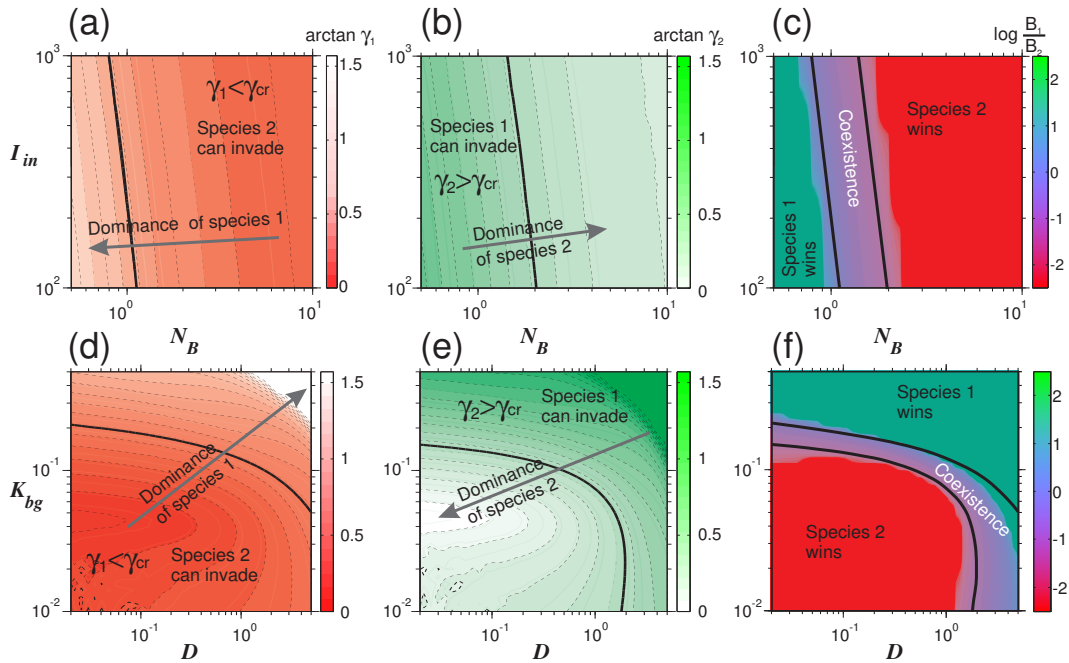}}
\caption{
Invasion analysis in the phytoplankton model. 
(Top panel) The slope of the invasion thresholds as a function of $N_B$ and $I_{in}$ numerically calculated for a monoculture of (a) species 1 and (b) species 2. The color intensity changes with $\arctan \gamma_i$ and characterizes the likelihood of invasion of (a) better light competitors (decreases with $\gamma$) and (b) better nutrient competitors (increases with $\gamma$). Black lines shows the boundary of the parameter range where the competitor can invade. (c) Compares the boundary of the coexistence range obtained from results for monocultures (Figs.~a and b) and from two-species modeling. The color shows the logarithm of the biomass ratio ($\log B_1/B_2$) after 20000 simulation days. 
(Bottom panel) The same analysis in dependence on $D$ and $K_{bg}$.
}
\label{fig:changeAll}
\end{center}
\end{figure*}

\subsection{Numerical simulations}
The value $\gamma$ quantifies the likelihood of invasion and can be used derive the competition outcome from the results obtained for each competitor alone. Consider species 1 (the better nutrient competitor) as the resident. 
Fig.~\ref{fig:changeAll}a shows the values of $\arctan \gamma_1$ in the $(I_{in}, N_B)$ plane, which were  numerically calculated from equilibrium resource distributions in the presence of this species. The intensity of red color decreases with $\gamma_1$ and characterizes the ability of a better light competitor (species 2) to invade.  Contrary, the dominance of species 1 over species 2 increases with $\gamma$. The black line shows the level $\gamma_1 = \gamma_{cr}$. In the region to the right of this line $\gamma_1 < \gamma_{cr}$ and therefore, species 2 can invade. 
In a similar manner, Fig.~\ref{fig:changeAll}b shows the slope of the invasion threshold, $\arctan \gamma_2$, in the presence of species 2 alone. Here, however, the color intensity increases with $\gamma_2$ and characterizes more favorable conditions for a better nutrient competitor. The black line is the boundary of the range $\gamma_2 > \gamma_{cr}$, where species 1 can invade. An intersection of these ranges shows the range of coexistence, in which  $\gamma_1< \gamma_{cr} < \gamma_2$ and each species can invade in the presence of its competitor. Fig.\ref{fig:changeAll}c compares this range, based on the one-species modeling (between black lines), with the range of coexistence obtained by two-species modeling (the blue area). 

This approach can be extended to include other model parameters. For instance, Figs.~\ref{fig:changeAll}d--f show that an increase of both the turbulent diffusivity, $D$, and background turbidity, $K_{bg}$ increases the slopes of the invasion thresholds and the competition outcome shifts from the dominance of species 2 through the range of coexistence to the dominance of species 1. In this figure again, the range of species coexistence predicted from simulation of one-species model (the area between two black lines in Fig.~\ref{fig:changeAll}f) is in a good agreement with the results obtained in two-species model (the blue area).    

\section{Discussion}

\subsection{Invasion thresholds}

In a uniform system a species can increase its biomass if its resource requirements are lower than the present level of ambient resources ($R^*$-rule). 
This rule, however, has to be generalized for spatially variable environments, where on the one hand the size of favorable area becomes a crucial factor, and on the other hand the resource heterogeneity provides an opportunity to compensate a lack of one resource by superfluous concentrations of another. 
To extend the competition theory to nonuniform systems Ryabov and Blasius (2011) recently suggested to replace the $R^*$-rule by the notion of an invasion threshold, which is defined
as the maximal requirements of successful invaders in the presences of a resident species in equilibrium. If the critical values $(N^*, I^*)$ of a species lie below the threshold, then the species can invade. 
In a nonuniform environment the shape of invasion thresholds can be complex and nonlinear. However, an approximated technique can be developed for competition between species with close resource requirements. For these species the invasion threshold can be approximated by a strait line on a double-logarithmic scale and the slope of this line (determined by the ratio of logarithmic resource gradients) becomes a critical determinant of the competition outcome.

The dependence of invasion thresholds on resource gradients, rather than on the local resource values, leads to new rules for invasion and coexistence. In particular a large value of $\gamma$ means that the invasion threshold approaches to a vertical line, which favors to good nutrient competitors (see Fig.~\ref{fig:comp}). By contrast, if $\gamma$ is small then the invasion thresholds is close to a horizontal line, therefore good light competitors are in more favorable conditions. 

This effect alters the mechanism of coexistence. 
In a uniform system two species can coexist if each of them mostly consumes its most limiting resource, and finally becomes self-limited by this resource. 
If however, the favorable area is bounded by resource availabilities, 
a somewhat opposite rule can be formulated: the monoculture of each species should shape resource distributions with a relatively smaller gradient of its most limiting resource. Then each competitor may benefit from its adaptation to a specific resource. 
As shown in (Ryabov and Blasius 2011), for phytoplankton competition this difference is even more striking, because $\gamma$ grows with light attenuation coefficient, $k$, and  decrease with nutrient content of cells, $\alpha$.
Thus, for coexistence in deep layers of a water column a low-light/high-nutrient adapted species should have a smaller value of $\alpha$, so that the nutrient is available also for species in the upper layers. Conversely, the species with high-light/low-nutrient adaptation should have a smaller coefficient $k$, thereby minimizing the light shading. Note that similar
correlations between consumption rates and resource adaptation were recently found in the experimental analysis of competition for light and phosphorus (Passarge et al. 2006). 

The dependence of competition outcome on environmental parameters in the phytoplankton model also reveals a number of intriguing results and shows that the intuition based on homogeneous models may fail in analysis of heterogeneous systems. For instance, Figs.~\ref{fig:changeAll}a-c show that an increase of both the incident light intensity, $I_{in}$, and nutrient concentrations at the bottom, $N_B$, decreases $\gamma$ and favors to the best light competitor (species 2). This is because both factors lead to a sharper biomass maximum and hence to a steeper nutrient gradient. Recall that in a uniform system an increase of $I$ would always favor to a good nutrient competitor, while an increase of $N$ would favor to a good light competitor. 

Further, Figs.~\ref{fig:changeAll}d-f show that an increase of the water turbidity, $K_{bg}$, or turbulent diffusivity, $D$, increases $\gamma$ and favors to the best nutrient competitor (species 1). Although leading to the same results, these changes involve different mechanisms: an increase of $K_{bg}$ increases the ratio $\gamma = c_I/c_N$ via an increase of the light gradient, $c_I$, whereas an increase of $D$ leads to the same result because the nutrient gradient $c_N$ becomes smaller.  

These results might have important outcomes for field research. 
For instance,  a decrease of $N_B$ or $D$ decreases  the nutrient availability in the water column, however, leads to opposite effects on the competition outcome.
Thus, a shift in the species composition caused by higher stratification of the ocean waters can be opposite to that caused by a reduction of nutrient levels in deep layers.  
This effect can possibly explain both positive and negative correlations between nutrient concentrations and the abundance of high-light adapted species observed along environmental gradients in the Atlantic Ocean (Johnson et al. 2006).

\subsection{Model assumptions}
To derive  the invasibility criterion (Eq.~\ref{eq:criteria}), an ``ideal'' system was considered. It was assumed that the resources are exponentially distributed, the biomass maximum is located in the  deep layer and $\mu_{\rm max, 1} = \mu_{\rm max, 2}$. Under these assumptions the analysis of competition becomes fairly simple, because the invasion threshold follows a straight line with slope $\gamma$ in double logarithmic space. 
Now consider a more general situation in which these  assumptions do not hold. 

First, if the resource distributions are not exponentially shaped then the invasion threshold takes the shape of a curve (blue solid line in Fig.~\ref{fig:model}c). 
However, a tangent line to this curve at the point $(N_1^*, \ I_1^*)$ is exactly the linearly approximated invasion threshold (blue dashed line).
Thus, if resources  deviate from exponential distributions, Eq.~\ref{eq:criteria} provides a first order approximation which is valid for species with close resource requirements. 
The larger the interval where resources change exponentially,  the larger the segment of the invasion threshold which follows the linear dependence in log-log scale.
Often the resource level changes exponentially for few and more orders of magnitude (Ryabov \& Blasius 2011). In this case the invasibility criterion Eq.~\ref{eq:criteria} is applicable, if differences in species half-saturation constants have the same or smaller order.

Second, it was assumed that the production layers are confined by the resource availability and located sufficiently far from the boundaries. The boundaries effect species distributions and survival conditions. For instance, an impenetrable boundary is a favorable factor for  species survival, because the cell diffusive flux reflects from the boundary and more cells can return into the production layer (Cantrell and Cosner 1991). 
The boundaries also influence the species spatial segregation. The species, which can occupy different depths in the deep layers, have to compete locally if the biomass maximum occurs in the benthic or surface layer, which strengthens the interspecific interactions. All these effects have a profound impact on species composition and can even reverse the outcome of competition, will be published elsewhere. 

Third, to focus on the role of resource gradients it was assumed that competitors have the same maximal growth rates, mortalities and dispersal.  In this settings an invasion threshold always passes through the resident's $(N^*, \ I^*)$ values. A more general approach (Ryabov and Blasius 2011) shows, however,  that if there is any difference in these parameters then the invasion threshold can be shifted towards higher or lower resource requirements. 
As a result an invader with, for instance, higher $\mu_{max}$ can invade even if it has higher requirements of both resources and two species can stably coexist via a positive correlation in the maximal growth rate and resource requirements (so-called gleaner-opportunist trade-off). 

\subsection{Comparison with other models}
There are few approaches suggested to describe phytoplankton competition in a water column. These approached can be classified based on the assumed intensity of water mixing.
Namely, Huisman and Weissing (1994, 1995) consider competition in well-mixed systems; by contrast, Yoshiyama et al. (2009) suggest an approach for poorly mixed environments; finally, Ryabov and Blasius (2011) and the present paper complement these studies for the case of moderate mixing.  Compare the main assumptions and outcomes of these models.

\begin{figure*}[tb]
\begin{center}
\centerline{\includegraphics[width=\textwidth]{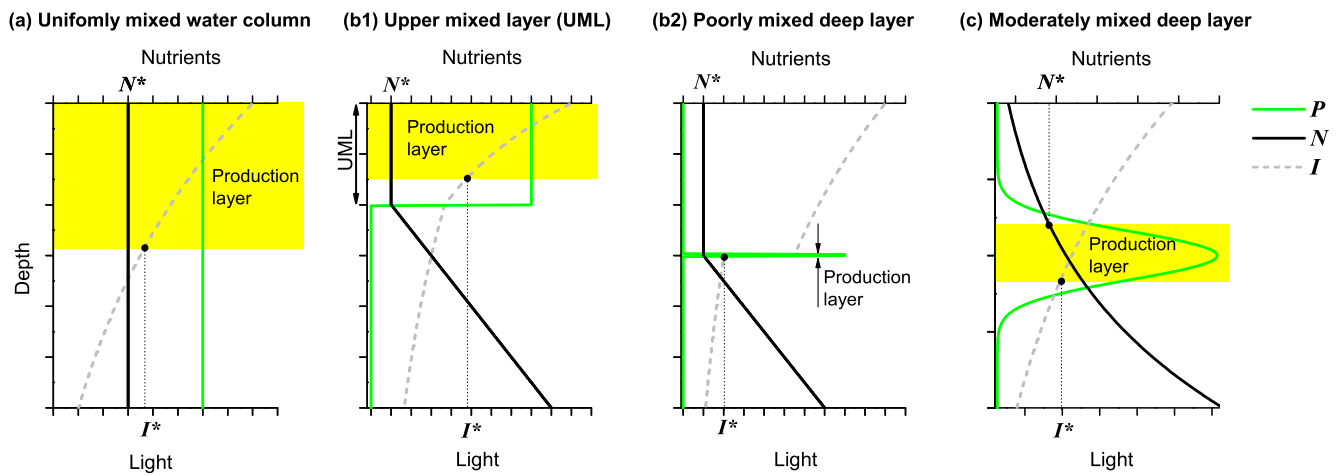}}
\caption{
Different modeling approaches for competition in a water column. 
(a) Huisman and Weissing (1994, 1995) assume an exponential decay of light and uniform distribution of nutrients and phytoplankton. The light limitation bounds the production layer (yellow area). 
 Yoshiyama et al. (2009) assume (b1) within the UML the biomass and nutrient are uniformly distributed and (b2) in the poorly mixed deep layer the biomass distribution is infinitely thin. The production layer is bounded by the availability of light in the UML and by both resources in the deep layer. (c) Ryabov and Blasius (2011) assume exponential resource distributions. The production layer is bounded by the availability of nutrients (upper boundary) and light (lower boundary) and has finite thickness. The biomass distribution can have an arbitrary shape. 
}
\label{fig:Models}
\end{center}
\end{figure*}

Huisman and Weissing (1994, 1995) performed an analysis, assuming that the light intensity decays exponentially, but phytoplankton biomass and nutrients are uniformly distributed (Fig.~\ref{fig:Models}a). The distributions of the competitors in this system completely overlap. The outcome of their competition changes with the ratio and absolute values of resource supplies. If competitors trade off in resource requirements, then (similar to Tilman 1980)  the outcome of competition depends on the resource ratio. If however, one competitor has a higher $\mu_{\rm max}$ then this species can benefit both from light and nutrients. Moreover, even having higher requirements for both resources, this species can win competition if the resource supplies are high enough.

Yoshiyama et al. (2009) consider competition in a stratified water column with a uniformly mixed upper layer and poorly mixed deep layer. For the upper layer the approach of Huisman and Weissing is applied, for the deep layer it is assumed that the biomass maximum is infinitely thin (Fig.~\ref{fig:Models}b). According to the last assumption, if species compete in the deep layer their distributions will never overlap. Thus, this model is more suitable if traits of two species are sufficiently different. For such species the model predicts that in the deep layer the resource supplies do not change the outcome of competition, they rather influence the depth of the biomass maxima. The outcome of competition changes, however, if the bulk biomass of at least one of the competitors occurs in the upper or benthic layer.

The approach of Ryabov and Blasius (2011) considers competition in the deep layer of a moderately or poorly mixed water column (Fig.~\ref{fig:Models}c). This approach is based on the comparison of the growth rate profiles in the presence of each competitor alone (Fig.~\ref{fig:GrShifts}). Consequently, the analysis does not depend on the biomass distribution, and the biomass maxima can overlap. 
Furthermore, this approach reveals a key role of resource gradients in community structuring.

A single mathematical model cannot present exactly the dynamics of a complex ecological system. All models contain some simplifications, and typically  a real system and model match qualitatively, but not quantitatively. In this sense, the invasion thresholds can become a useful tool of the qualitative analysis. In the model considered here the invasion thresholds have a simple linear shape in a log-log scale. In another model this shape can be non-linear or linear, but in another scale. If, however,  this shape and its dependence on system parameters can be deduced numerically or analytically, than we can also project shifts in the species composition for this system. 
Thus, this analysis can be further extended for a wide spectrum of spatially heterogeneous models, in which other biotic or abiotic factors, such as gradients of temperature, predation, mortality, etc., are taken into account.

\section{Acknowledgments}
I thank B.Blasius for comments and useful suggestions.

\begin{table*}[p]
\begin{center}
\caption{Parameters values and their meaning}
\begin{tabular}{llll}
\hline
\hline \\[-2ex]
     Symbol & Interpretation  &  Units & Value  \\ [0.5ex]
     \hline\\[-2ex]
     \multicolumn{4}{l}{\textbf{Independent variables}}\\
        $t$ & Time & h & - \\ [0.5ex]
$z$ & Depth & m & - \\ [0.5ex]
\hline\\[-2ex]
\multicolumn{4}{l}{\textbf{Dependent variables}}\\
$P(z, t)$ & Population density & cells m$^{-3}$ &  \\ [0.5ex]
$I(z, t)$ & Light intensity & $\mu$mol photons m$^{-2}$ s$^{-1}$ &  \\[0.5ex]
$N(z, t)$ & Nutrient concentration & mmol nutrient m$^{-3}$ &  \\[0.5ex]
\hline\\[-2ex]
\multicolumn{4}{l}{\textbf{Parameters}}\\
$I_{\rm in}$ & Incident light intensity & $\mu$mol photons m$^{-2}$ s$^{-1}$ & 600 (100 - 1000) \\[0.5ex]
$K_{\rm bg}$ & Background turbidity    & m$^{-1}$ & 0.1 \\ [0.5ex]
$Z_B$  & Depth of the water column     & m & 100 \\ [0.5ex]
$N_B$  & Nutrient concentration at $Z_B$ & mmol nutrient m$^{-3}$ & 2 (0.1-10) \\[0.5ex]
$D$  & Vertical turbulent diffusivity& cm$^2$ s$^{-1}$ & 0.3 (0.02 - 5) \\ [0.5ex]
$\mu_{\rm max}$ & Maximum specific growth rate     & h$^{-1}$ & 0.04 \\ [0.5ex]
$m$  & Specific loss rate        & h$^{-1}$ & 0.01 \\[0.5ex]
\multicolumn{4}{l}{\textbf{Species 1 -- best nutrient competitor}}\\
$H_{I, 1}$  & Half-saturation constant, light limitation & $\mu$mol photons m$^{-2}$ \ s$^{-1}$ & 20 \\[0.5ex]
$H_{N, 1}$  & Half-saturation constant, nutrient limitation  & mmol nutrient m$^{-3}$ &  0.04\\[0.5ex]
$\alpha_1$ & Nutrient content of phytoplankton & mmol nutrient cell$^{-1}$ & 8 $\times$10$^{-10}$ \\[0.5ex]
$k_1$  & Absorption coefficient of a phytoplankton cell & m$^2$ cell$^{-1}$ & 6$ \times$10$^{-10}$ \\[0.5ex]
\multicolumn{4}{l}{\textbf{Species 2 -- best light competitor}}\\
$H_{I, 2}$  & Half-saturation,  light limitation & $\mu$mol photons m$^{-2}$ \ s$^{-1}$ & 15 \\[0.5ex]
$H_{N, 2}$  & Half-saturation constant  nutrient limitation & mmol nutrient m$^{-3}$ &   0.065\\[0.5ex]
$\alpha_2$ & Nutrient content of phytoplankton & mmol nutrient cell$^{-1}$ & 5 $\times$10$^{-10}$ \\[0.5ex]
$k_2$  & Absorption coefficient of a phytoplankton cell & m$^2$ cell$^{-1}$ & 6$ \times$10$^{-10}$ \\[0.5ex]
     \hline
\end{tabular}
\label{tab:1}
\end{center}
\end{table*}

\end{document}